\documentclass[AER]{AEA}

\usepackage{longtable}
\usepackage{comment}
\usepackage{natbib}
 \usepackage{graphicx}
\usepackage[backref=page]{hyperref}
\hypersetup{
colorlinks = true,
linkcolor = cyan,
citecolor = magenta,
urlcolor = blue,
}

\draftSpacing{1.5}

\begin{document}

\title{The Nobel Family}
\shortTitle{The Nobel Family}
\author{Richard S.J. Tol\thanks{Tol: Department of Economics, University of Sussex, BN1 9SL Falmer, United Kingdom, r.tol@sussex.ac.uk; Institute for Environmental Studies, Vrije Universiteit Amsterdam, The Netherlands; Department of Spatial Economics, Vrije Universiteit Amsterdam, The Netherlands; Tinbergen Institute, Amsterdam, The Netherlands; CESifo, Munich, Germany; Payne Institute of Public Policy, Colorado School of Mines, Golden, CO, USA; College of Business, Abu Dhabi University, Abu Dhabi, United Arab Emirates.}}
\date{\today}
\pubMonth{February}
\pubYear{2023}
\pubVolume{}
\pubIssue{}

\begin{abstract}
Nobel laureates cluster together. 696 of the 727 winners of the Nobel Prize in physics, chemistry, medicine, and  economics belong to one single academic family tree. 668 trace their ancestry to Emmanuel Stupanus, 228 to Lord Rayleigh (physics, 1904). Craig Mello (medicine, 2006) counts 51 Nobelists among his ancestors. Chemistry laureates have the most Nobel ancestors and descendants, economics laureates the fewest. Chemistry is the central discipline. Its Nobelists have trained and are trained by Nobelists in other fields. Nobelists in physics (medicine) have trained (by) others. Economics stands apart. Openness to other disciplines is the same in recent and earlier times. The familial concentration of Nobelists is lower now than it used to be.
\end{abstract}

\maketitle

\section{Introduction}
The Nobel Prize is the highest accolade in academia. Who are the winners? What made them into what they are? This paper sheds partial light on that last question, mapping the academic ancestry of Nobelists. There are 727 Nobel laureates. There are 25 family trees with a single Nobelist, 4 trees with 2 Nobelists, and 1 tree with 696 Nobelists. This is a remarkable agglomeration of excellence.

The clustering of Nobel Prize winners has been documented before \citep{Zuckerman1996, Chan2015}, but not in terms of academic genealogy. The only comparable effort is limited to the winners of the Sveriges Riksbank Prize in Economic Sciences in Memory of Alfred Nobel \citep{Tol2022ehjet}. The current paper extends that family tree to the Nobel Prizes in physics, chemistry, and medicine or physiology. (The prizes for literature and peace are of an entirely different nature.)

A family tree shows more than just clustering. It allows for the identification of key figures in research training as revealed by the number of and closeness to Nobel descendants. It also distinguishes Nobelists who are insiders from those who are not. The paper also uses a newly defined measure of cross-closeness \citep{Tol2023nobelware} to identify Nobelists who studied with other Nobelists. I also analyze differences between the four disciplines in terms of their respective concentration of Nobelists and their openness to other disciplines.

The paper proceeds as follows. Section \ref{sc:data} discusses the data and methods. Section \ref{sc:results} shows the results for Nobel descendants, ancestors, and peers, as well as differences between disciplines and changes over time. Section \ref{sc:conclude} concludes.

\section{Data and methods}
\label{sc:data}

\subsection{Data}
I constructed the \emph{academic} ancestry of all Nobel laureates, focusing on PhD advisor-advisee relations in recent times and on wider mentor-mentee relations for earlier periods.\footnote{There are intriguing familial relationships as well, with fathers and sons, mother and daughter, brothers, and brothers-in-law all winning Nobel prizes.} The main source of information is \href{https://academictree.org/physics/tree.php?pid=36889}{AcademicTree}. The database was largely complete at the start of this project and updated where needed.

The AcademicTree is a Wiki. For recent times, its main source of information is ProQuest, a database of all PhD theses completed at a consortium of major research universities. A number of volunteers have added great historical depth to the data.\footnote{There is occasional mythical depth too. Tracing Isaac Newton's ancestry to William of Ockham is one thing, relating Ockham to Jesus Christ is something else.} Other volunteers have added data about themselves or people close to them. The result is uneven coverage. Prominent researchers, however, are likely to be included.

I added Nobel laureates and their ancestors who were not already included using \href{https://www.genealogy.math.ndsu.nodak.edu/}{Mathematics Genealogy}, \href{https://genealogy.repec.org/}{RePEc Genealogy}, Wikipedia and a range of other sources, including biographies, obituaries, and PhD theses. In a few cases, I emailed individuals.\footnote{All except one responded. People are eager to talk about their mentors.}

The definition of ``advisor'' is problematic. Formalities and practice vary strongly over time, between countries, between disciplines, and between institutions. It is not uncommon among prominent emeriti in Western Europe to have only a Master's degree and in the generations before that, we find people who were home-schooled or self-taught. In other places or recent times, a PhD counts for little; it is the \textit{Habilitation} that matters, or the second PhD, or the post-doctoral fellowship. In some universities, professors jealously guard their students whereas in other places it takes a village to train a researcher. On top of that, the formal advisor may differ from the actual teacher. These caveats notwithstanding, this is the best data available.

Ancestors were added until the respective Nobelists were connected to the main family. If no connection was possible, four generations of ancestors were added, if known. The resulting tree has 33 generations, with Erasmus as \textit{Urahn}.

\subsection{Methods}
Data were transferred to Matlab and stored as a directed acyclic graph or polytree for analysis and visualization. Representation as a polytree offers a number of standard measures of centrality. I use the harmonic mean distance, where distance is the number of edges between two nodes. The harmonic mean is defined for unconnected polytrees, as is the case here, and emphasizes proximate over distant relations. I define distance as the distance to a Nobel laureate, rather than to any node. Besides the standard outcloseness for academic ancestors and incloseness for descendants, I also define and use crosscloseness to measure the distance to Nobel siblings and cousins. I analyze these measures for all Nobel laureates and separately for Physics, Chemistry, Physiology or Medicine, and Economics.

More precisely, the distance from a node $i$ in a graph to the rest of this graph can be measured by the H\"{o}lder mean
\begin{equation}
\label{eq:holder}
    D_{i}(h) = \left (\frac{1}{|J|} \sum_{j \in J} D_{j,i}^h \right )^\frac{1}{h}
\end{equation}
where $D_{j,i}$ is the distance from node $i$ to any node $j$, that is, the number of edges between node $i$ and node $j$. The set $J$ typically includes all nodes $j \neq i$ but may be restricted to nodes with a particular characteristic. Here, $J$ contains only Nobelists.

For $h=1$, the H\"{o}lder mean is the arithmetic mean. This can be computed using the Matlab function \textsc{centrality}, which is included in the standard release. Note that $D_{i}(1) = \infty$ unless node $i$ descends from \emph{all} other nodes in set $J$. This makes it less suitable for any application to unconnected graphs, as is the case here.

For $h=-1$, the H\"{o}lder mean is the harmonic mean, which is bounded if some nodes in the network cannot be reached. In other words, the harmonic mean applies to connected as well as unconnected subgraphs: For unreachable nodes $D_{j,i} = \infty$ so $1/D_{j,i} = 0$. \citet{Marchiori2000} propose this as a measure of distance, \citet{GilSchmidt1996} its inverse as a measure of closeness.

The H\"{o}lder mean distance can be used to emphasize proximity at the expense of distal relationships. Close relations are further emphasized as $h$ becomes more negative.

Equation (\ref{eq:holder}) is an \emph{outcloseness} measure. Outcloseness on a polytree measures ancestry. Replacing $D_{j,i}$ by $D_{i,j}$ in Equation (\ref{eq:holder}) yields an \emph{incloseness} measure, measuring descent.

Outcloseness and incloseness measure the vertical distance, between parents and children. The horizontal distance, crosscloseness \citep{Tol2023nobelware}, is of interest too\textemdash siblings can be just as influential as parents. The horizontal distance of node $i$ to $j$ on a polytree is defined as
\begin{equation}
    H_{i,j}(n) = \frac{| D_{k,i} = D_{k,j} = n |}{\max(|D_{k,i} = n|,|D_{k,j} = n| )}
\end{equation}
That is, distance equals the number of shared ancestors of generation $n$ divided by the maximum number of ancestors. In biology, $H_{i,j}(1) = 1$ for siblings, $H_{i,j}(1) = 0.5$ for half-siblings, and $H_{i,j}(1) = 0$ for everyone else. $H(i,j)(2) = 0.5$ for first cousins, $H(i,j)(3) = 0.25$ for second cousins, and so on.

Having constructed the matrix $H$ of horizontal distances, the inverse of the generalized mean of Equation (\ref{eq:holder}) then defines crosscloseness. 

\section{Results}
\label{sc:results}
Figure \ref{fig:main} shows the main family tree of 696 Nobel laureates. Figure \ref{fig:all} in the Appendix shows all trees, Table \ref{tab:loners} lists the Nobel prize winners who are not part of the main tree. Nobelists are colour-coded by discipline. Node size is proportional to the sum of out-, in-, and crosscloseness. Figure \ref{fig:main} shows a thick cluster of nodes, with some separation between physics, chemistry, and medicine, with economics as an outgrowth.

There are 360 professor-student pairs who both won the Nobel Prize, 255 in the same discipline. These numbers increase to 863, 431 in the same discipline, if we include grandprofessor-grandstudent pairs and more distant relationships. This highlights just how tightly knit the Nobel tree is.

\subsection{Nobel descendants}
\href{https://neurotree.org/neurotree/tree.php?pid=620}{Emmanuel Stupanus}\footnote{Individuals mentioned are linked to their profile on AcademicTree.} is the nearest common ancestor of 668 Nobelists, almost all of the 696 Nobelists in the main tree. Stupanus was a 17th-century professor at the University of Basel, best known for his opposition to empirical evidence in medicine. He trained a few students\textemdash \href{https://neurotree.org/neurotree/tree.php?pid=619}{Franz de le Bo\"{e}}, \href{https://neurotree.org/neurotree/tree.php?pid=25691}{Johann Bauhin} and \href{https://neurotree.org/neurotree/tree.php?pid=25692}{Nikolaus Eglinger}\textemdash but their students were more numerous and influential. See Figures \ref{fig:stupanus2strutt}, \ref{fig:stupanus2baeyer} and \ref{fig:bauhin}.

The Nobelist with the most Nobel descendants (228) is \href{https://neurotree.org/neurotree/tree.php?pid=14817}{John Strutt, Lord Rayleigh} (physics, 1904). His student, \href{https://academictree.org/physics/tree.php?pid=13139}{Joseph Thompson} (physics, 1906) comes second, with 227 Nobelists. Seven other Nobelists have more than 100 Nobel descendants: \href{https://academictree.org/chemistry/tree.php?pid=21401}{Adolf von Baeyer} (chemistry, 1905), \href{https://academictree.org/chemistry/tree.php?pid=5484}{Wilhelm Ostwald} (chemistry, 1909), \href{https://academictree.org/chemistry/tree.php?pid=13140}{Ernest Rutherford} (chemistry, 1908)), \href{https://academictree.org/chemistry/tree.php?pid=22091}{Emil Fischer} (chemistry, 1902), \href{https://academictree.org/physics/tree.php?pid=1942}{Max Born} (physics, 1954), \href{https://academictree.org/physics/tree.php?pid=1943}{Niels Bohr} (physics, 1922), and \href{https://academictree.org/chemistry/tree.php?pid=6989}{Walther Nernst} (chemistry, 1920). Five of these hold Nobel Prizes in chemistry, four in physics.

\href{https://neurotree.org/neurotree/tree.php?pid=14817}{John Strutt} is the Nobelist with the most descendants (126) who won the Nobel Prize in physics. \href{https://academictree.org/chemistry/tree.php?pid=21401}{Adolf von Baeyer} tops the list in chemistry, with 107 Nobel descendants. Strutt and von Baeyer descend from \href{https://neurotree.org/neurotree/tree.php?pid=619}{de le Bo\"{e}}; see Figures \ref{fig:stupanus2strutt} and \ref{fig:stupanus2baeyer}. The numbers are much lower in medicine: \href{https://academictree.org/chemistry/tree.php?pid=22102}{Otto Warburg} (1931) has the largest Nobel descent at 35. The prize in economics is much younger. \href{https://academictree.org/econ/tree.php?pid=12124}{Wassily Leontief} (1973) has the largest number of Nobel descendants (15).

\href{https://academictree.org/meteorology/tree.php?pid=66786}{Georg Lichtenberg} is the central-most professor in the network. Lichtenberg was an 18th century physicist at the University of G\"{o}ttingen, best known for his work on electricity. He also trained a large number of scientists, who in turn trained more. See Figure \ref{fig:lichtenberg} for the first two generations.  In both Lichtenberg and Stupanus, we find a common ancestor who is not renowned for his contributions to science, but who was influential in training young scientists, including in the art of training young researchers.

The central-most \emph{Nobel} professor, and the 12th-most central professor, is \href{https://neurotree.org/neurotree/tree.php?pid=14817}{John Strutt}. \href{https://academictree.org/physics/tree.php?pid=13140}{Ernest Rutherford} is the highest-ranked Nobelist (joint 75th) in chemistry, \href{https://academictree.org/chemistry/tree.php?pid=22102}{Otto Warburg} in medicine (479th), \href{https://academictree.org/econ/tree.php?pid=12124}{Wassily Leontief} in economics (595th).

\subsection{Nobel ancestry}
\href{https://neurotree.org/neurotree/tree.php?pid=47322}{Craig Mello} (medicine, 2006) has the most Nobel ancestry: 51 of his academic ancestors won the Nobel Prize. \href{https://academictree.org/chemistry/tree.php?pid=92631}{Georges Kohler} (medicine, 1984) comes second with 42, followed by \href{https://academictree.org/chemistry/tree.php?pid=2848}{Robert Horvitz} (medicine, 2002) with 31 and \href{https://academictree.org/chemistry/tree.php?pid=5082}{Arthur Kornberg} (chemistry, 2006) and \href{https://neurotree.org/neurotree/tree.php?pid=2536}{David Julius} with 30 (medicine, 2021). Four of the top five won in medicine, seven of the top 10; the rest is in chemistry. The physics Nobelist with the Noblest ancestry is \href{https://academictree.org/physics/tree.php?pid=94211}{Eric Cornell} (2001) with 23, ranking 14th. \href{https://academictree.org/econ/tree.php?pid=616505}{Esther Duflo} (2018) the highest ranked economist, a shared 134th, with 8 Nobel ancestors.

The central-most student is \href{https://neurotree.org/neurotree/tree.php?pid=58070}{Victor Ambros} who was \href{https://neurotree.org/neurotree/tree.php?pid=47322}{Craig Mello}'s professor and therefore closer to Mello's academic ancestors. Mello is the most-central Nobel student and the 3rd-most central student, after \href{https://academictree.org/chemistry/tree.php?pid=92629}{Fritz Melchers}, who was one of \href{https://academictree.org/chemistry/tree.php?pid=92631}{Georges Kohler}'s professors. Seven of the top ten Nobelists are in medicine, three in chemistry. \href{https://academictree.org/physics/tree.php?pid=144428}{Martin Perl} (1995) is the highest-ranked physicist at 29, \href{https://academictree.org/econ/tree.php?pid=616505}{Esther Duflo} the highest ranked economist at 82.

As noted above, 31 of the 727 Nobelists are not connected to main family. There are 66 Nobelists who have no Nobel ancestry and no Nobel peers. Another 130 Nobelists have fellow students who won the Nobel Prize but no professors who did.

\subsection{Shared ancestry}
\begin{figure}[hp]
    \centering
    \includegraphics[width=\textwidth]{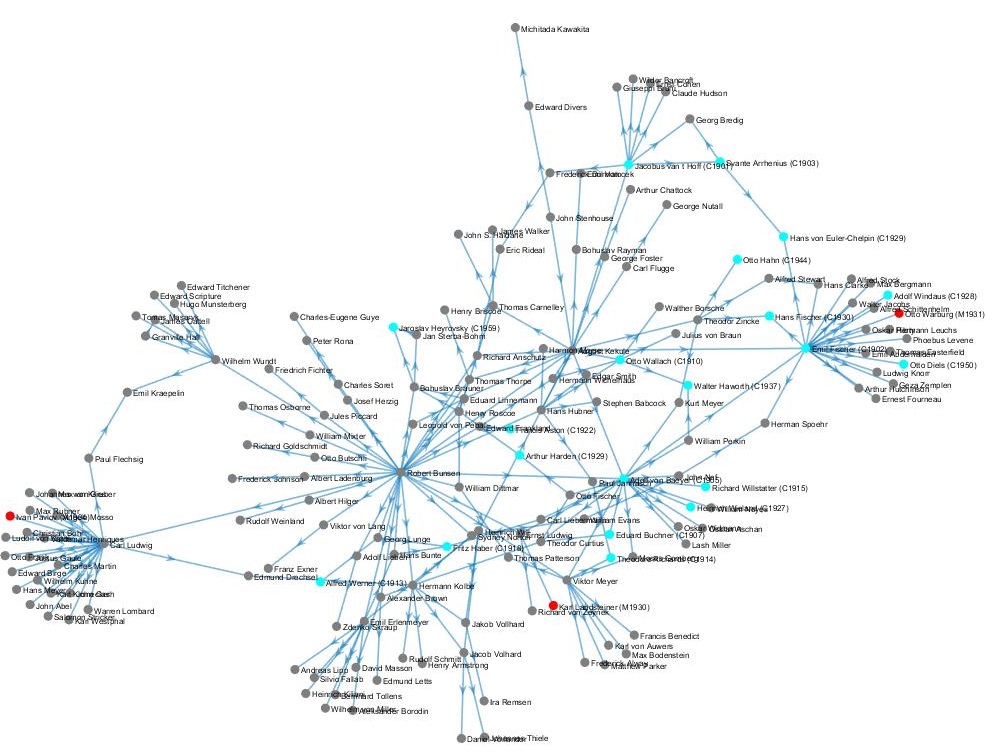}
    \caption{Academic siblings and cousins of Emil Fischer. The colour denotes the discipline: red = medicine, blue = chemistry, grey = not a Nobel laureate, but an ancestor of Nobelists.}
    \label{fig:fischer}
\end{figure}

The central-most fellow student of Nobelists is \href{https://academictree.org/chemistry/tree.php?pid=22091}{Emil Fischer} (chemistry, 1902) who, with \href{https://academictree.org/chemistry/tree.php?pid=21404}{August Kekul\'{e}} and \href{https://academictree.org/chemistry/tree.php?pid=21401}{Adolf von Baeyer} as professors, studied with an amazing cast of later Nobelists. Figure \ref{fig:fischer} shows all grandstudents of Fischers' grandprofessors\textemdash that is, his academic siblings and cousins\textemdash who either won the Nobel prize or have descendants who did. This is a remarkable cluster of excellence.

\href{https://academictree.org/chemistry/tree.php?pid=27672}{Harold Urey} (chemistry, 1934) is the 2nd-most central peer. He studied under \href{https://academictree.org/chemistry/tree.php?pid=49964}{Gilbert Lewis} and \href{https://academictree.org/physics/tree.php?pid=1943}{Niels Bohr}, together with many other prominent scholars. The top 12 central-most enNobeled fellow students are all chemists. \href{https://neurotree.org/neurotree/tree.php?pid=23353}{Karl Landsteiner} (1928) is the highest-ranked Nobelist in medicine at 13. \href{https://academictree.org/physics/tree.php?pid=36900}{Julian Schwinger} (1965) tops the physics list at 17, \href{https://academictree.org/econ/tree.php?pid=23347}{Tjalling Koopmans} (1975) the economics list at 68, although he has more academic cousins in physics than in economics. 

\subsection{Differences between disciplines}
Figure \ref{fig:main} and the results above suggests that different disciplines play different roles. This is underlined in Table \ref{tab:proximate} (proximal descent) and Table \ref{tab:distal} (distal descent). Table \ref{tab:proximate} shows that 96 Nobel laureates in chemistry have students who won the Nobel prize, 66 in chemistry, 12 in physics, and 18 in medicine. Medicine laureates trained chemistry laureates but no physics ones. Economics laureates neither trained nor were trained by laureates in other disciplines. Table \ref{tab:distal} reveals a similar pattern, with chemistry firmly in the centre, training more of the laureates in other disciplines and receiving more training from them. Some physics laureates can trace their ancestry to medicine ones. Some economics laureates have ancestry in physics and chemistry, or in medicine.

Table \ref{tab:ancdesc} amplifies this result. The average Nobelist has 4.6 Nobel ancestors\textemdash therefore, the average Nobelist also has 4.6 Nobel descendants. These numbers vary between fields. Chemistry Nobelists have the most Nobel ancestors (5.9), economics Nobelists the fewest (1.0). This difference is statistically significant, as are the differences with in-between physics (4.7) and medicine (4.9). On average, physics (3.5) and chemistry (3.5) have the most Nobel ancestors from their own field, followed by medicine (1.9) and economics (0.8). The majority (59\%) of Nobel ancestors of Nobel laureates in medicine are from other fields, about a third (34\%) and a fifth (21\%) for chemistry and physics, and only 6\% for economics. These differences are statistically significant.

Table \ref{tab:ancdesc} also shows the average number of descendants. Chemistry (7.0) and physics (6.2) Nobelists have the most Nobel descendants, followed by medicine (2.6) and economics (0.8). The number of Nobel descendants \emph{by field} equals the number of Nobel ancestors \emph{by field}. Medicine laureates have the largest share (43\%) of Nobel descendants in other fields, statistically significantly more than physics (29\%) and medicine (22\%). Economics laureates have no Nobel descendants in other fields.

Overall, clustering of Nobel laureates in family trees is strongest in chemistry and physics, and weakest in economics. Chemistry laureates train most laureates in other fields; medicine laureates are trained most by laureates in other fields. Economics is the most isolated of the four fields.

\subsection{Changes over time}
\begin{figure}[hp]
    \centering
    \includegraphics[width=\textwidth]{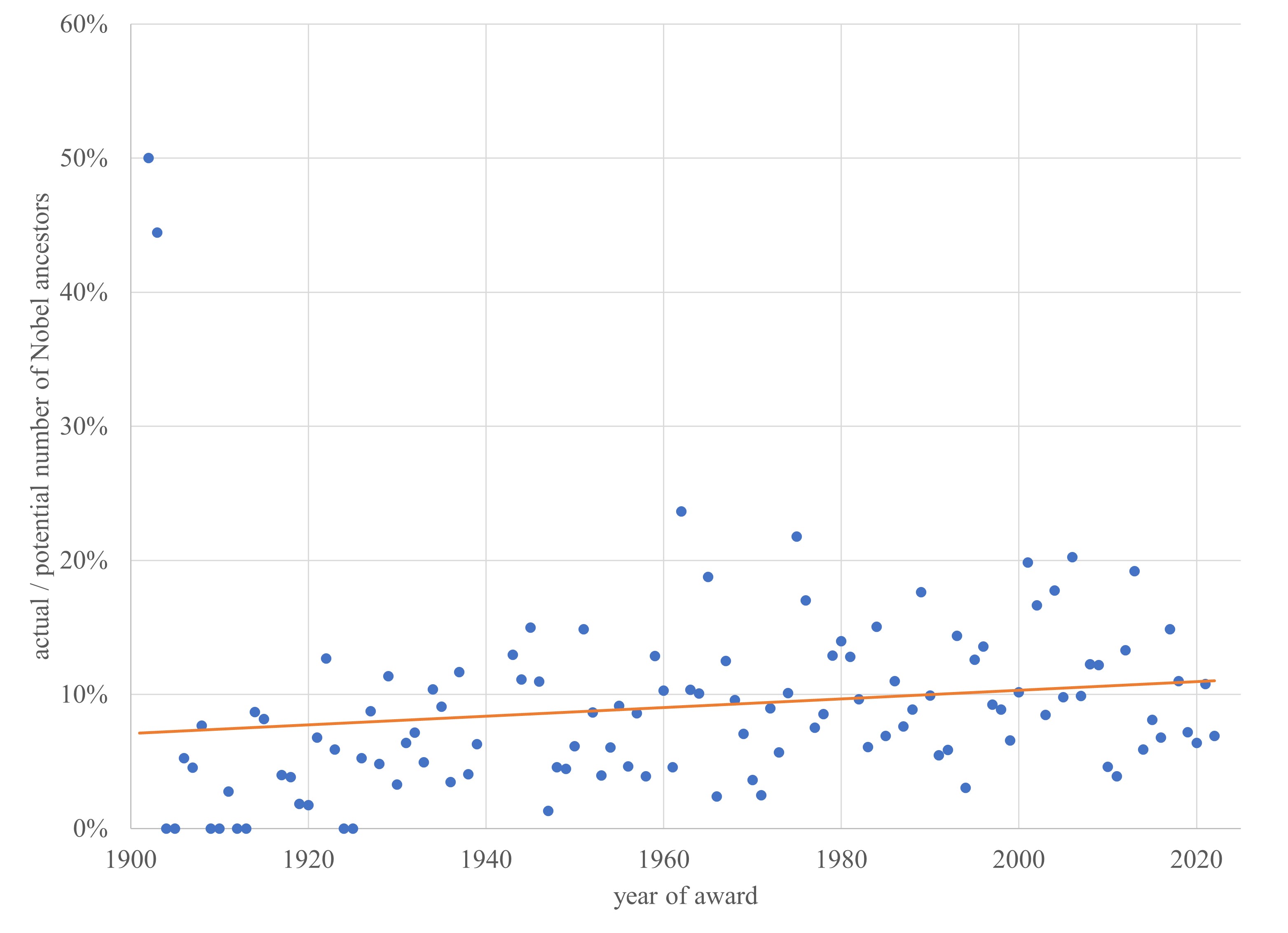}
    \caption{The number of Nobel ancestors over the number  of Nobel laureates over time.}
    \label{fig:ancestime}
\end{figure}

Figure \ref{fig:ancestime} plots the number of Nobel ancestors divided by the number of Nobel laureates against the year of the award. There is a slight upward trend. That is, the number of Nobel ancestors of Nobel laureates has grown faster than the number of Nobel laureates.

\begin{figure}[hp]
    \centering
    \includegraphics[width=\textwidth]{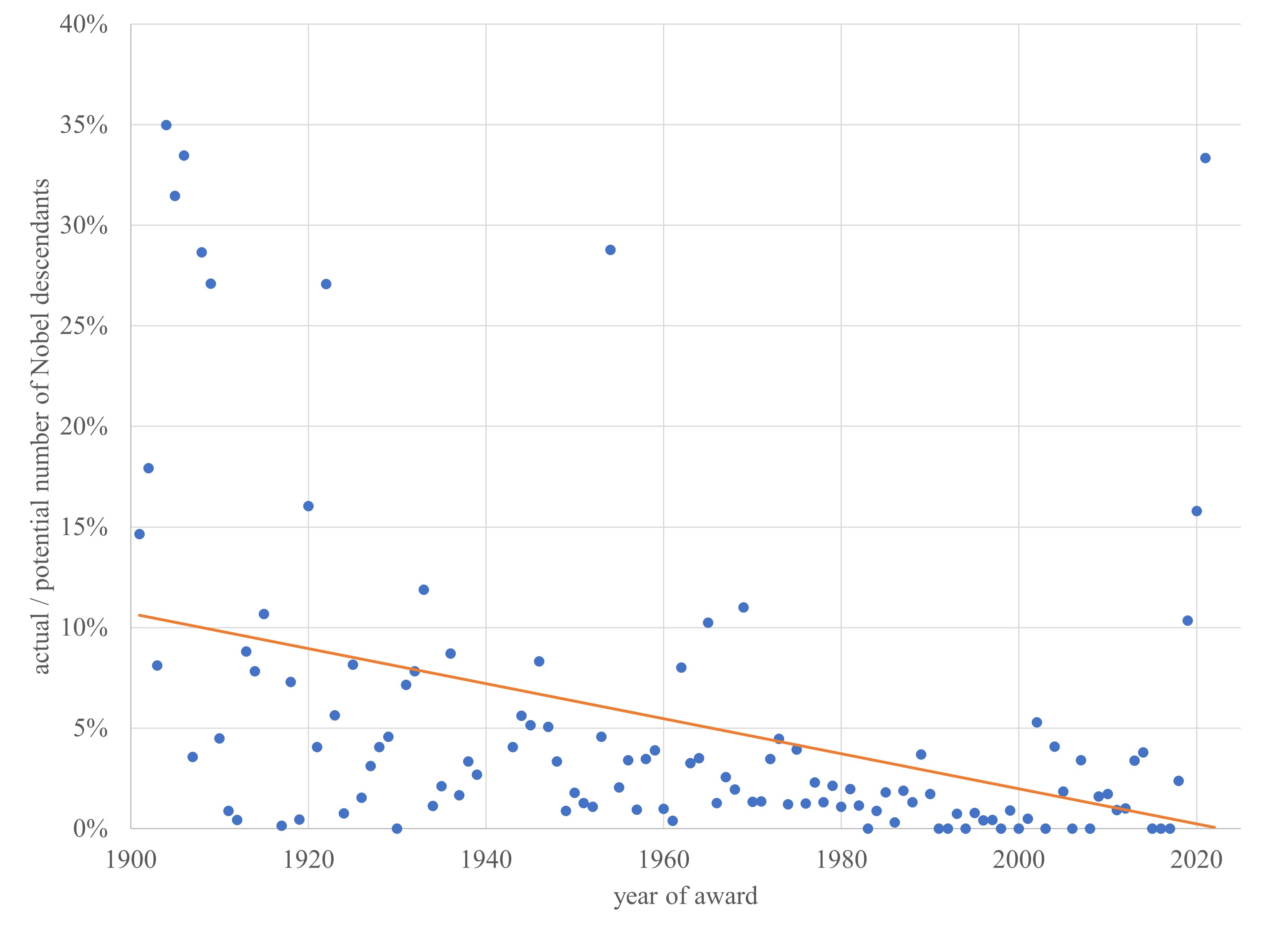}
    \caption{The number of Nobel descendants over the number of Nobel laureates over time.}
    \label{fig:descentime}
\end{figure}

Figure \ref{fig:descentime} plots the number of Nobel descendants divided by the number of Nobel laureates against the year of the award. There is a clear downward trend. That is, the number of Nobel laureates has grown faster than the number of Nobel descendants of Nobel laureates. The slight upward trend in Figure \ref{fig:ancestime} notwithstanding, the Nobel tree has grown less concentrated over time.

Figures \ref{fig:ancestors} and \ref{fig:descendants} plot the fraction of Nobel ancestors and descendants, respectively, of Nobel laureates who won in a different field against the year of the award. There has been no significant or substantial change over time. Overall, fields are as open (or closed) to outside influence now as they were in the past.

Figure \ref{fig:noanc} plots the fraction of Nobel laureates who do not have a Nobel prize winner among their ancestors. This fraction starts relatively high. The early Nobelists studied with venerable researchers who could not have won a prize that had yet to be instituted. From around 1950 onwards, however, the fraction is roughly stable, even though the number of past Nobelists keeps increasing.

Figure \ref{fig:noties} plots the fraction of Nobel laureates neither whose professors nor whose fellow students won the Nobel prize. This fraction has increased over the last 40 years or so. As with Figure \ref{fig:noties}, this suggests that the Nobel prize has opened up to people of non-Nobel families.

\section{Discussion and conclusion}
\label{sc:conclude}
I construct the academic family tree of all 727 winners of the Nobel Prize in physics, chemistry, and medicine and the Nobel Memorial Prize in economics. 96\% of all laureates belong to one family tree; 92\% of laureates are related in the sense that their professor's professor's ... professor was Emmanuel Stupanus. 31\% of Nobel prize winners descend from Lord Rayleigh, who won the physics prize in 1904. 7\% of Nobel laureates are ancestors of Craig Mello, who won the medicine prize in 2006. Chemistry (economics) laureates have the highest (lowest) number of Nobelists among their ancestors and descendants. Chemistry Nobelists have trained and are trained by Nobelists in other fields. Physics Nobelists have trained others, and medicine laureates are trained by others. Economics sits largely apart. Openness to other disciplines has not changed over time, but the familial concentration of Nobelists has fallen.

The analysis in this paper is limited to formal teaching relationships. It does not include other forms of scientific collaboration, such as co-authorship \citep{Kademani2005, Fields2015a, Fields2015b, Bai2021}, informal mentoring, collegiality, and competition. Such relationships are important too, but harder to map. I do not look at the almae matres of the Nobelists or where they did their most important work \citep{Schlagberger2016}. I study neither the methods and flow of ideas \citep{Chan2015}\textemdash indeed, Emmanuel Stupanus would be aghast at the empirical research of most of his Nobel descendants\textemdash nor citations \citep{Bjork2014, Sangwal2015, Zhang2019swan, Frey2020, Kosmulski2020}.

A key question is not answered in this paper. Is the concentration of Nobelists because the best professors select the best students \citep{Athey2007} and teach them well \citet{Jones2021}, or is it because Nobelists have a strong voice in later awards and disproportionally nominate their proteges \citep{Zuckerman1996}? Examination of the minutes of the awarding committees suggests that the latter explanation is at least partially true \citep[][but see \citep{Tol2022nobelcand}]{Economist2021}. Further study would be welcome.

\bibliographystyle{aea}
\bibliography{master}

\newpage
\begin{table}[h]
    \centering
    \begin{tabular}{l | r r r r r r}
 & \multicolumn{6}{c}{students} \\
professors & Physics & Chemistry & Medicine & Economics & Any	& None \\ \hline
Physics	& 98 & 16 & 2 & 0 & 116 & 107\\
Chemistry &	12 & 66 & 18 & 0 & 96 & 92 \\
Medicine & 	0 & 15 & 91 & 0 & 106 & 119 \\
Economics &	0 & 0 & 0 & 42 & 42 & 49 \\ \hline
    \end{tabular}
    \caption{Nobel laureates as PhD advisors.}
    \label{tab:proximate}
\end{table}

\begin{table}[h]
    \centering
    \begin{tabular}{l | r r r r r r}
 & \multicolumn{6}{c}{descendants} \\
ancestors & Physics & Chemistry & Medicine & Economics & Any	& None \\ \hline
Physics	& 165 & 68 & 46 & 3 & 282 & 136\\
Chemistry &	113 & 145 & 92 & 3 & 353 & 116 \\
Medicine & 	17 & 43 & 120 & 1 & 181 & 141 \\
Economics &	0 & 0 & 0 & 47 & 47 & 58 \\ \hline
    \end{tabular}
    \caption{Nobel laureates as academic ancestors.}
    \label{tab:distal}
\end{table}

\begin{table}[]
    \centering
    \begin{tabular}{l r r r r r} \hline
	&	Any	&	Physics	&	Chemistry	&	Medicine	&	Economics	\\ \hline
Nobel ancestors	&	4.60	&	4.65	&	5.90	&	4.93	&	0.98	\\
	&	(0.22)	&	(0.32)	&	(0.44)	&	(0.51)	&	(0.15)	\\
Nobel ancestors from own field	&		&	3.52	&	3.54	&	1.92	&	0.84	\\
	&		&	(0.24)	&	(0.25)	&	(0.20)	&	(0.12)	\\
Fraction from other fields	&		&	0.21	&	0.34	&	0.59	&	0.06	\\
	&		&	(0.02)	&	(0.03)	&	(0.03)	&	(0.03)	\\ \hline
Nobel descendants	&	4.60	&	6.17	&	6.96	&	2.59	&	0.84	\\
	&	(0.73)	&	(1.66)	&	(1.91)	&	(0.43)	&	(0.20)	\\
Nobel descendants in own field	&		&	3.52	&	3.54	&	1.92	&	0.84	\\
	&		&	(0.92)	&	(0.93)	&	(0.32)	&	(0.20)	\\
Fraction in other fields	&		&	0.29	&	0.43	&	0.22	&	0.00	\\
	&		&	(0.04)	&	(0.04)	&	(0.03)	&	(0.00)	\\ \hline
\end{tabular}
    \caption{Average (standard error) number of Nobel ancestors and descendants of Nobelists}
    \label{tab:ancdesc}
\end{table}

\begin{figure}[hp]
    \centering
    \includegraphics[width=\textwidth]{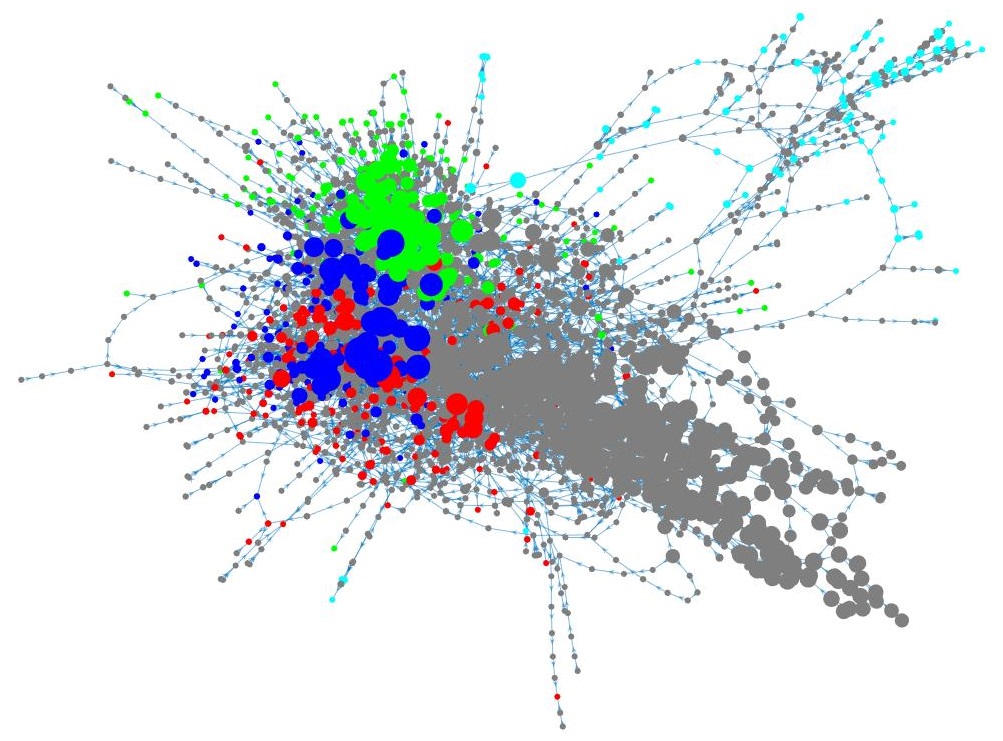}
    \caption{The main Nobel network. The colour denotes the discipline: red = medicine, blue = physics, green = chemistry, light blue = economics, grey = not a Nobel laureate. The size denotes proximity, the sum of in-, out- and cross-closeness, to Nobel laureates.}
    \label{fig:main}
\end{figure}

\newpage \appendix 
\setcounter{page}{1}
\renewcommand{\thepage}{A\arabic{page}}
\setcounter{table}{0}
\renewcommand{\thetable}{A\arabic{table}}
\setcounter{figure}{0}
\renewcommand{\thefigure}{A\arabic{figure}}
\setcounter{equation}{0}
\renewcommand{\theequation}{A\arabic{equation}}

\section{Additional results}
\begin{table}[h]
    \centering
    \begin{tabular}{l c c} \hline
    name & field & year \\ \hline
    \multicolumn{3}{c}{Nobel professor-student pairs with no known ancestry}\\
    John McLeod \& Frederick Banting & medicine & 1923 \\
    Ragnar Frisch \& Trygve Haavelmo  & economics & 1969 \& 1989  \\ 
    Maurice Allais \& Gerard Debrue & economics & 1983 \& 1988 \\
    Isamo Nakasuki \& Hiroshi Amano & physics & 2014 \\
    \multicolumn{3}{c}{Nobelists with known ancestry}\\
    Gustaf Dalen & physics & 1912 \\
    Ronald Ross & medicine & 1920 \\
    William Murphy & medicine & 1934 \\
    Alexander Fleming & medicine & 1945 \\
    Andre Cournand & medicine & 1956 \\
    Gertrude Elion & medicine & 1988 \\
    Jens Skou & chemisty & 1997 \\
    Zhores Alferov & physics & 2000 \\
    Christopher Pissarides & economics & 2010 \\
    Dan Shechtman & chemistry & 2011 \\
    Youyou Tu & medicine & 2015 \\
    Michael Houghton & medicine & 2020 \\
    \multicolumn{3}{c}{Nobelists with no known ancestry}\\
    Niels Finsen & medicine & 1903 \\
    Antonio Muniz & medicine & 1949 \\
    Leo Esaki & physics & 1973 \\
    Godfrey Hounsfield & medicine & 1979 \\
    Jack Kilby & physics & 2000 \\
    Hideki Shirakawa & chemistry & 2000 \\
    Koichi Tanaka & chemistry & 2002 \\
    Yves Chauvin & chemistry & 2005 \\
    Barry Marshall & medicine & 2005 \\
    Robin Warren & medicine & 2005 \\
    Shuji Nokamura & physics & 2014 \\
    Peter Ratcliffe & medicine & 2019 \\
    Syokuro Manabe & physics & 2021 \\
    \hline
    \end{tabular}
    \caption{Nobelists who are not connected to the main tree}
    \label{tab:loners}
\end{table}

\begin{figure}[h]
    \centering
    \includegraphics[width=\textwidth]{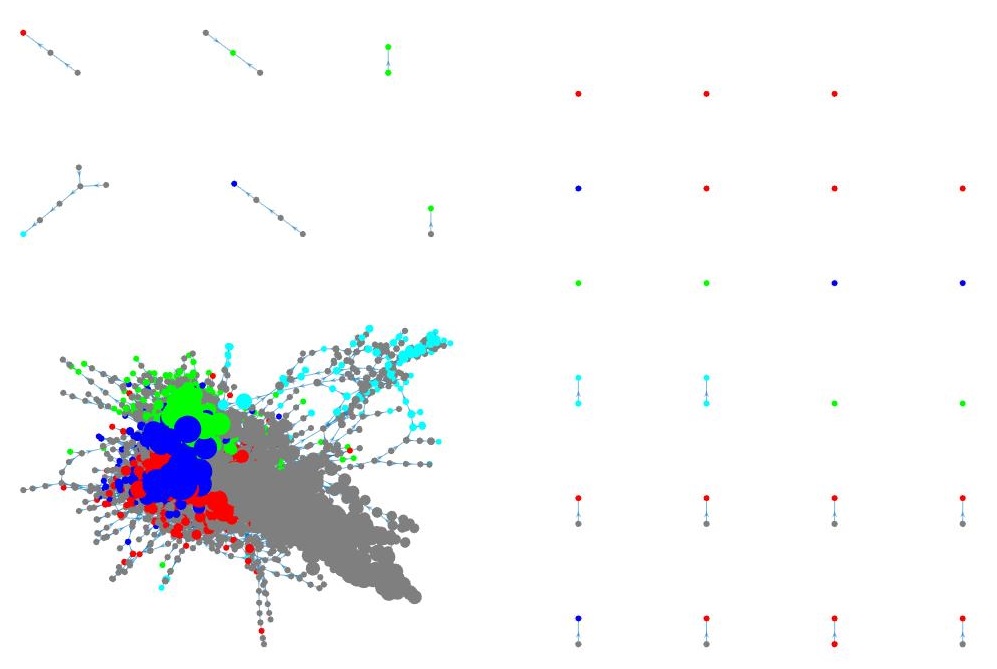}
    \caption{All Nobel networks. The colour denotes the discipline: red = medicine, blue = physics, green = chemistry, light blue = economics, grey = not a Nobel laureate. The size denotes proximity, the sum of in-, out- and cross-closeness, to Nobel laureates.}
    \label{fig:all}
\end{figure}

\begin{figure}[h]
    \centering
    \includegraphics[width=\textwidth]{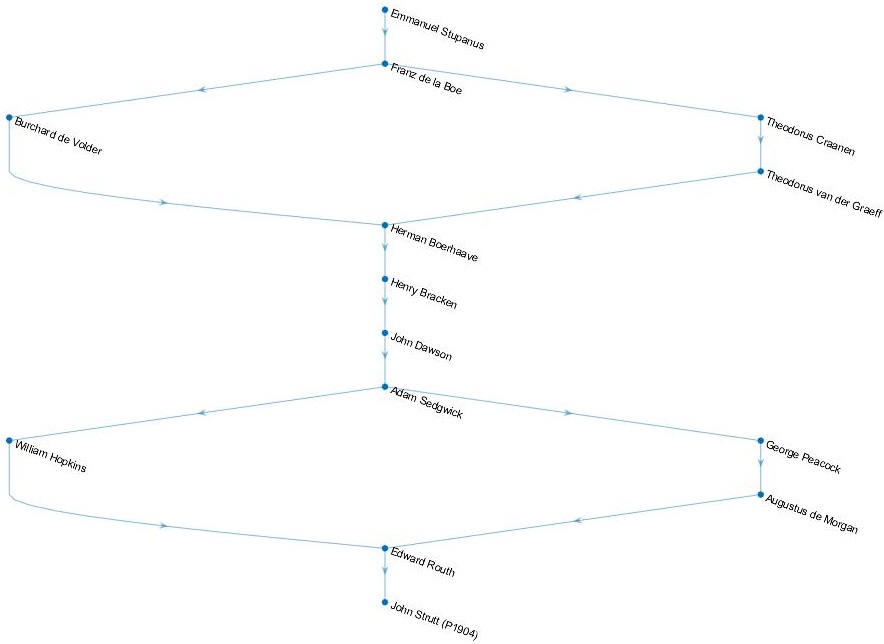}
    \caption{Strutt's descent from Stupanus.}
    \label{fig:stupanus2strutt}
\end{figure}

\begin{figure}[h]
    \centering
    \includegraphics[width=\textwidth]{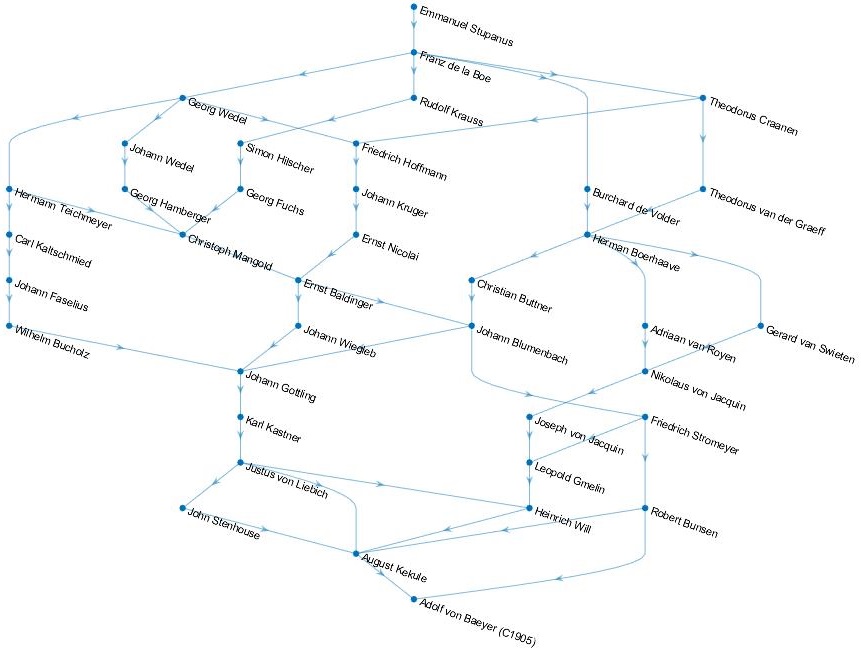}
    \caption{Baeyer's descent from Stupanus.}
    \label{fig:stupanus2baeyer}
\end{figure}

\begin{figure}[h]
    \centering
    \includegraphics[width=\textwidth]{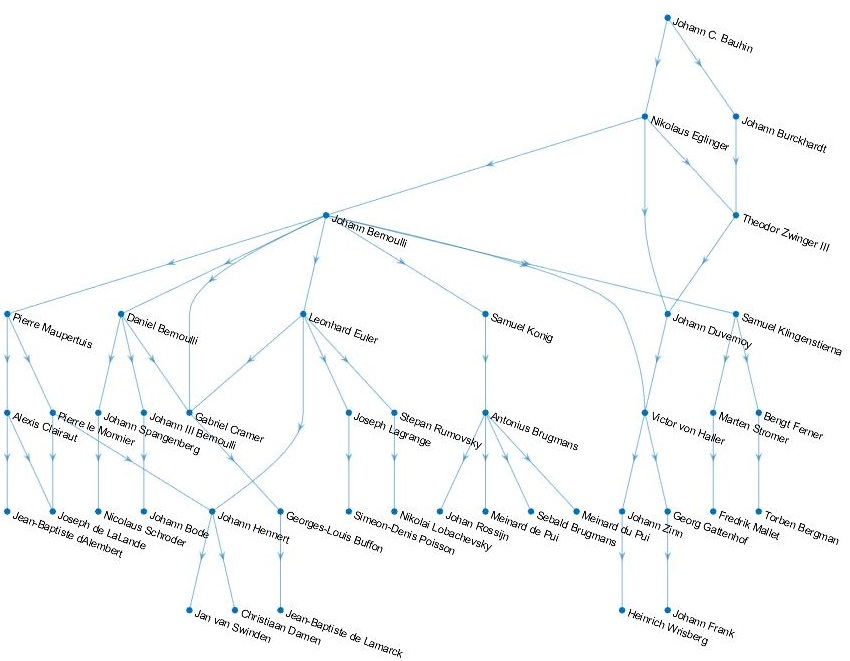}
    \caption{Five generations of descendants of Johann Bauhin.}
    \label{fig:bauhin}
\end{figure}

\begin{figure}[h]
    \centering
    \includegraphics[width=\textwidth]{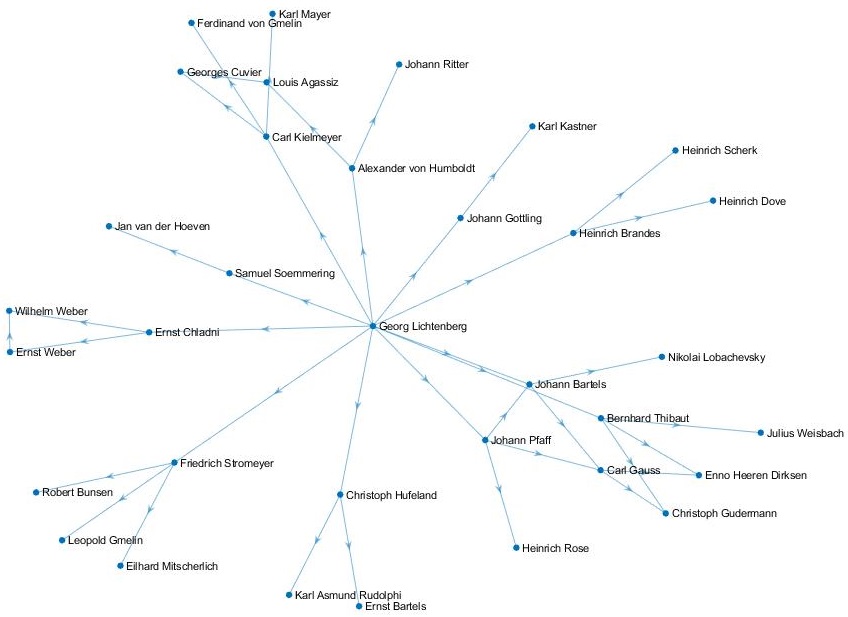}
    \caption{Students and grand-students of Georg Christoph Lichtenberg, the central-most scholar.}
    \label{fig:lichtenberg}
\end{figure}

\begin{figure}[h]
    \centering
    \includegraphics[width=\textwidth]{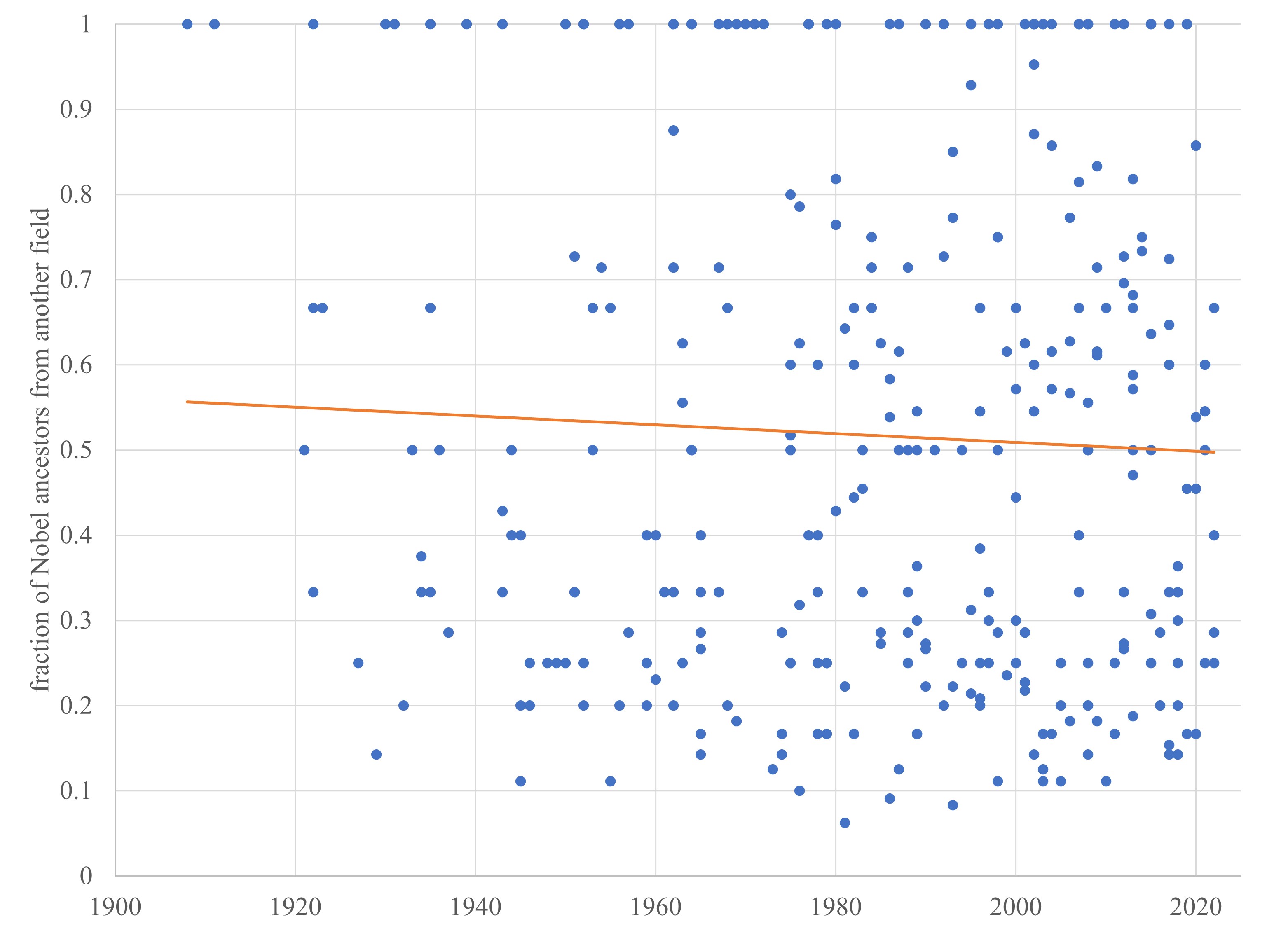}
    \caption{The fraction of Nobel ancestors of Nobelists who one their Nobel Prize in a different field over time.}
    \label{fig:ancestors}
\end{figure}

\begin{figure}[h]
    \centering
    \includegraphics[width=\textwidth]{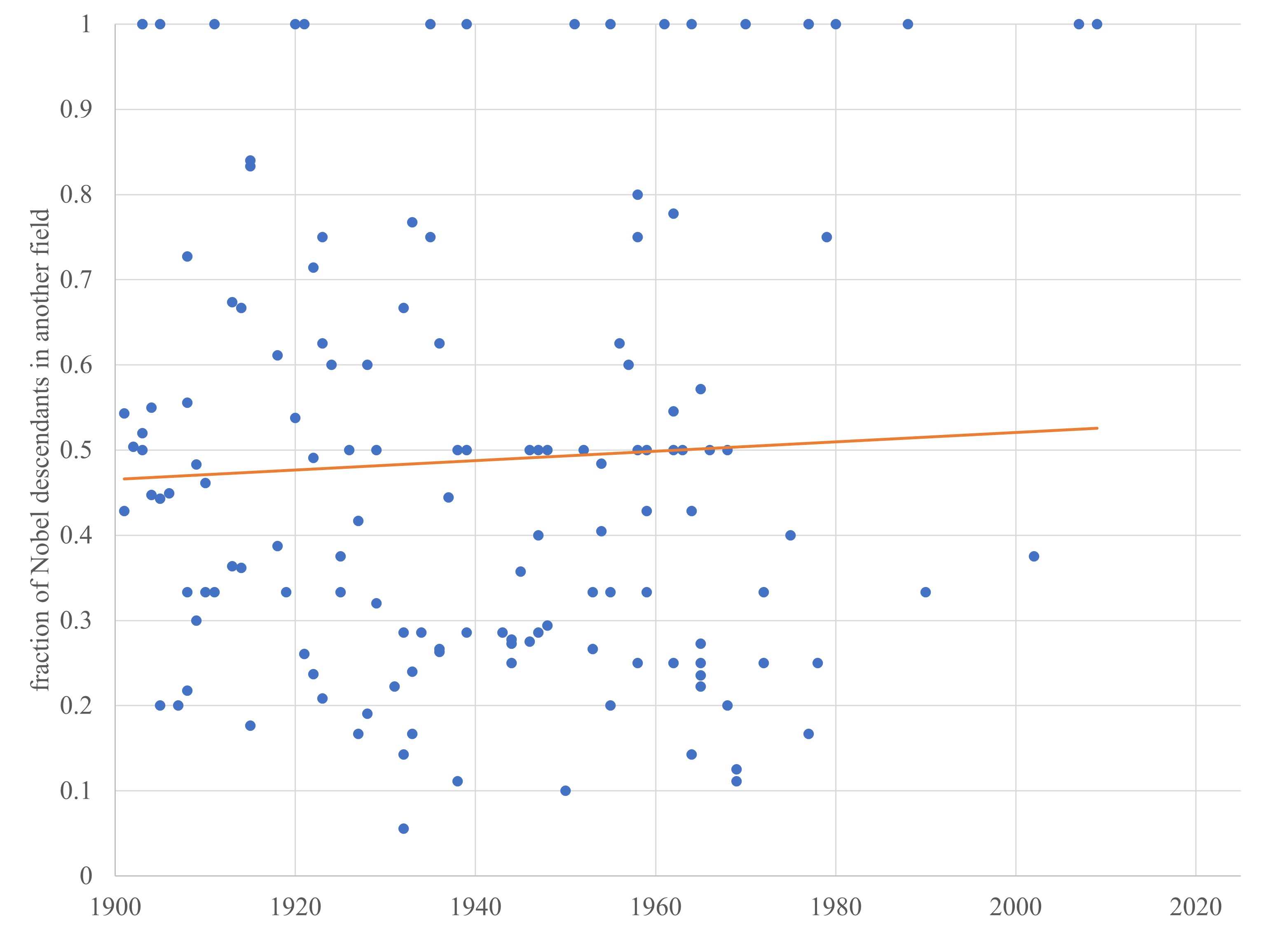}
    \caption{The fraction of Nobel descendants of Nobelists who one their Nobel Prize in a different field over time.}
    \label{fig:descendants}
\end{figure}

\begin{figure}[h]
    \centering
    \includegraphics[width=\textwidth]{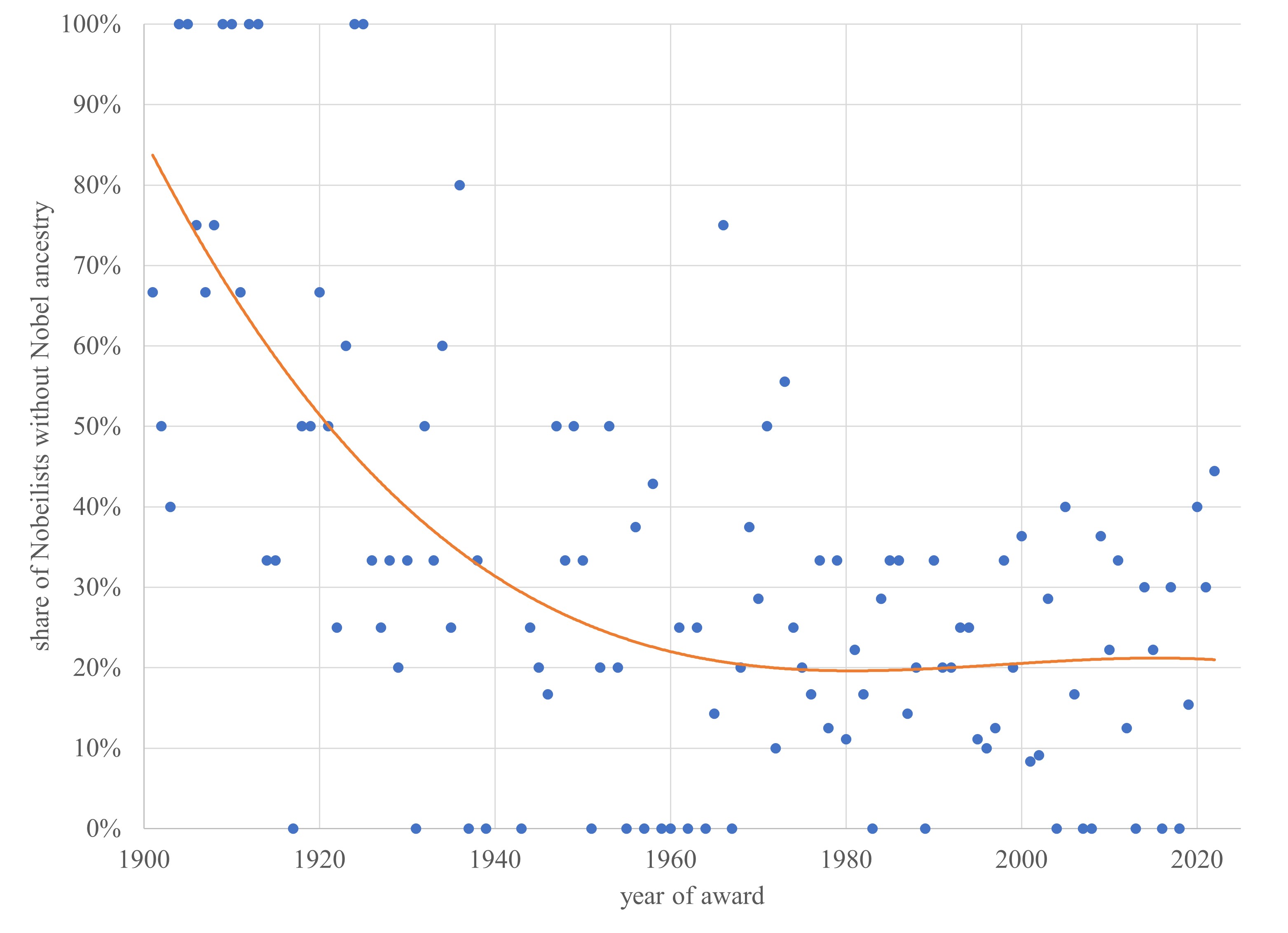}
    \caption{The fraction of Nobelists who have no Nobel ancestry over time.}
    \label{fig:noanc}
\end{figure}

\begin{figure}[h]
    \centering
    \includegraphics[width=\textwidth]{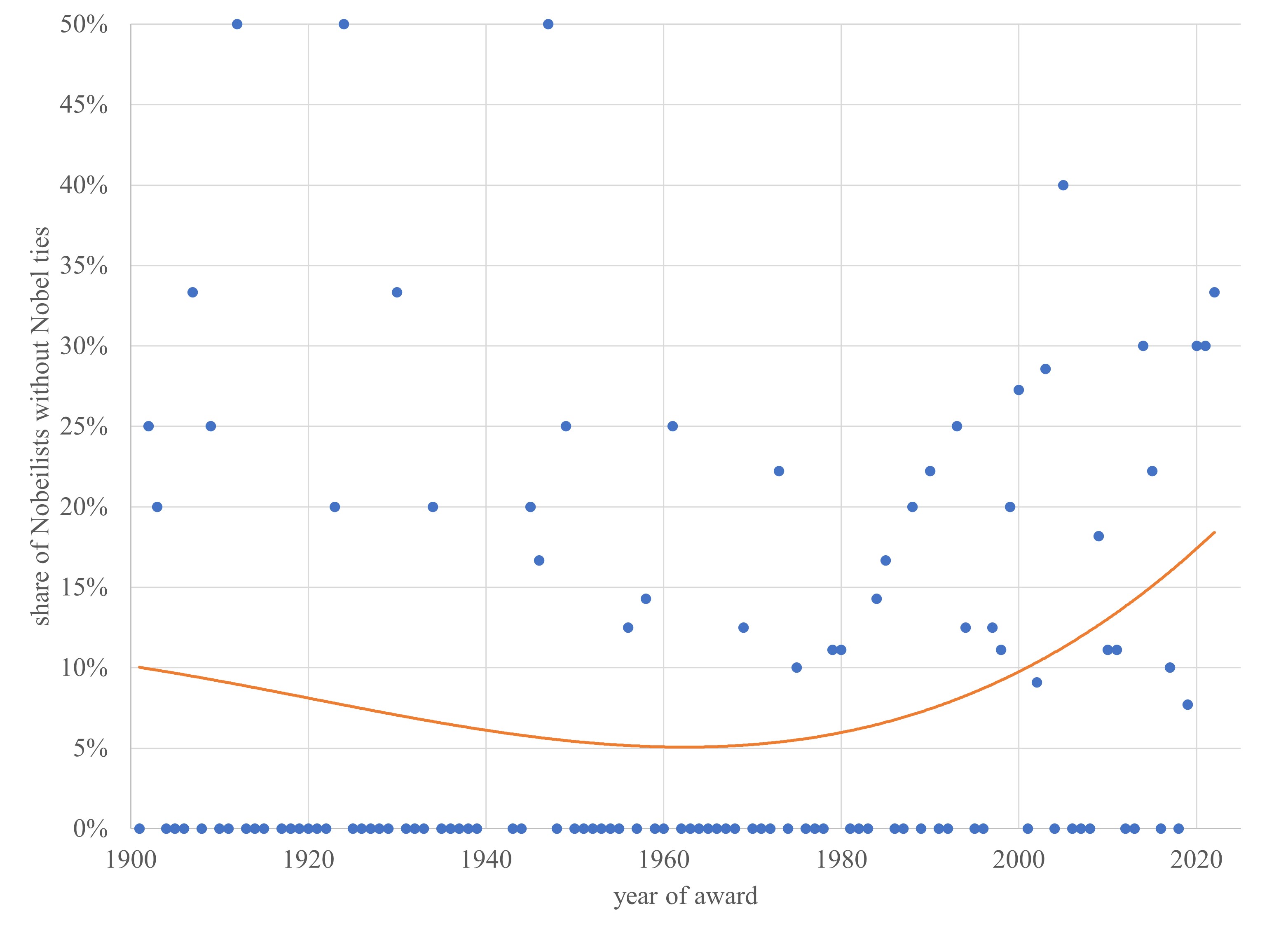}
    \caption{The fraction of Nobelists who have no Nobel ancestry or peers over time.}
    \label{fig:noties}
\end{figure}

\end{document}